\documentclass[twocolumn,showpacs,preprintnumbers,amssymb]{revtex4-1}
\usepackage{graphicx}
\usepackage{dcolumn}
\usepackage{color}
\usepackage{bm}

\def \L{\Lambda}
\def \e{\epsilon}

\def \d{\delta}
\def \k{\kappa}
\def \s{\sqrt}
\def \be{\begin{equation}}
\def \ee{\end{equation}}
\def \ben{\begin{eqnarray}}
\def \een{\end{eqnarray}}
\def \o{\omega}
\def \O{\Omega}

\begin{document}

\title{Thermodynamics for the k-essence Emergent Reissner-Nordstrom-de Sitter Spacetime}

\author{Goutam Manna }
\altaffiliation{goutammanna.pkc@gmail.com}
\affiliation{Department of Physics, Prabhat Kumar College, Contai, Purba Medinipur-721404, India}

\author{Bivash Majumder}
\altaffiliation{bivashmajumder@gmail.com}
\affiliation{Department of Mathematics, Prabhat Kumar College, Contai, Purba Medinipur-721404, India}

\author{Ashoke Das}
\altaffiliation{ashoke.avik@gmail.com}
\affiliation{Department of Mathematics, Raiganj University, Raiganj, West Bengal-733134, India.}

\begin{abstract}

The {\bf k-}essence emergent Reissner-Nordstrom-de Sitter spacetime has exactly mapped on to the Robinson-Trautman (RT) type spacetime with cosmological constant $\L$ for certain configuration of {\bf k-}essence scalar field. Theoretically, we evaluated that the thermodynamical quantities for the RT type emergent black hole is different from the usual one in the presence of kinetic energy of the {\bf k-}essence scalar field i.e.,  the dark energy density.  We restrict ourselves into the fact that the dark energy density (K) is to be unity, then the effective temperature and pressure both are negative for the RT type emergent black hole which implies that the system is thermodynamically unstable when the charge $Q\neq 0$ and the emergent spacetime is only dark energy dominated  and it does not radiate when $Q=0$. The thermodynamically unstable situation is physically plausible only when we consider spin degrees of freedom of a system. We have made this analysis in the context of dark energy in an emergent gravity scenario having {\bf k-}essence scalar fields $\phi$ with a Dirac-Born-Infeld type lagrangian. The scalar field also satisfies the emergent equation of motion at $r\rightarrow\infty$. 

\end{abstract}

\maketitle

\section{Introduction}
\label{intro}
The differences between the {\bf k-}essence theory with non-canonical kinetic terms and the relativistic field theories with canonical kinetic terms are that nontrivial dynamical solutions of the {\bf k-}essence equation of motion not only spontaneously break Lorentz invariance but also, change the metric for the perturbations around these solutions. Thus the perturbations propagate in the so-called {\it emergent or analogue} curved spacetime \cite{babi1,babi2,babi3,babi4,babi5} with the metric which is different from the gravitational one.  The emergent gravity metric $\tilde G_{\mu\nu}$ is not conformally equivalent to the gravitational metric $g_{\mu\nu}$. The general form of the lagrangian for {\bf k-}essence model is: $L=-V(\phi)F(X)$ where $\phi$ is {\bf k-}essence scalar field, $X=\frac{1}{2}g^{\mu\nu}\nabla_{\mu}\phi\nabla_{\nu}\phi$ and does not depend explicitly on $\phi$ to start with \cite{babi1,babi2,babi3,babi4,babi5,gm1,gm2,gm3,scherrer1,scherrer2,scherrer3}. The Lagrangian for {\bf k-}essence scalar fields contain non-canonical kinetic terms. Relevant kinds of literature  \cite{gor1}-\cite{gor15} for such fields discuss issues on cosmology, inflation, dark matter, dark energy and strings.

Research interest in the context of the Hawking radiation as well as the thermodynamics of black holes has gained momentum in several years. 
Based on a specific
Dirac-Born-Infeld \cite{born1,born2,born3} model for the {\bf k}-essence scalar field and specific standard physical black hole spacetimes such as Schwarzschild, Reissner-Nordstrom, the emergent spacetime is establishing conformal invariance with the Barriola-Velenkin, Robinson-Trautman (RT) type spacetimes \cite{gm1,gm2} and for the  Kerr, Kerr-Newman and Kerr-Newman-AdS background, the emergent metrics are satisfying Einstein equation for large $r$ along $\theta=0$ \cite{gm2,gm3}. Also, in the above references, the authors have established the modification of the Hawking temperature \cite{haw1}-\cite{haw15} in the presence of dark energy in an emergent gravity scenario.

Researchers
\cite{cai,zhao1,zhao2,dolan,seki,gibbon1,gibbon2,ren,urano,li,kubiznak} focus on the thermodynamics of black holes with cosmological constant $\L$ and  describe thermodynamical first law of black hole horizon spacetimes with $\L$, through the mechanical first law (MFL). The MFL is a geometrical relation and it relates the variations of the mass parameter, horizon area and other supplemental quantities. For example, the MFL of the Schwarzschild black hole is equivalent to its thermodynamical first law by using the mass as the only variable which describes the effect of a single horizon on the MFL.
The mass parameter $M$ and the cosmological constant $\L$ are regarded as two independent variables in the MFL for spacetimes 
with a cosmological constant $\L$ which possess the black hole event horizon (BEH) and cosmological event horizon (CEH).
Also, in \cite{buri,ghosh}, the authors have studied the thermodynamics and phase structure of the de Rham, Gabadadze, and Tolley (dRGT) massive
gravity \cite{rham1,rham2} spacetime. They have evaluated the mass, temperature and entropy of the dRGT black hole, and also performed thermodynamical stability analysis. They have shown that the presence of the graviton mass completely changes the black hole thermodynamics. They also have discussed how the linear term $\gamma r$ in the metric, which is the unique character of the dRGT massive gravity, affected the structure of the spacetime.

In this work, we investigate theoretically the thermodynamical behaviour of a black hole in the presence of dark energy in an emergent gravity scenario for the Reissner-Nordstrom-de Sitter background using the {\bf k-}essence model. Mention that one class of theoretical model to investigate the effects of the presence of dark energy on cosmological scenarios are {\bf k-}essence models where background metrics are the usual gravitational metric.  We find here the emergent spacetime in the presence of dark energy can be mapped on to the Robinson-Trautman(RT) type spacetime with cosmological constant $\L$. Also, we find when the charge $Q=0$ and the dark energy density is constrained to be unity then Hawking temperature is zero and {\it when the charge $Q\neq 0$, the temperature is negative i.e. this RT type of emergent spacetime is thermodynamically unstable. So we can say that the thermodynamics of this  RT type black hole is different from \cite{gm2}.} Note that in the Ref. \cite{gm2}, when the dark energy density is unity this black hole has zero Hawking temperature also but the charge $Q$ is non-zero. Also, we clearly mention that in this work we can not study the massive gravity in the context of the emergent gravity scenario.

The paper is organized as follows:
we are briefly describe the {\bf k-}essence theory and emergent gravity \cite{babi1,babi2,babi3,babi4,babi5} where $\tilde G_{\mu\nu}$ contains the dark energy field $\phi$ and this should satisfy the emergent gravity equations of motion in section 2. 
Again, for  $\tilde G_{\mu\nu}$ to be a blackhole metric, it has to satisfy the Einstein field equations. In section 3, we have shown that the emergent gravity metric having background metric is Reissner-Nordstrom-de Sitter (RNdS) mapping on to Robinson-Trautman (RT) type metric with cosmological constant $\L$. So Einstein field equation is automatically satisfied. The {\bf k-}essence scalar field $\phi$ is also satisfied emergent gravity equation of motion at $r\rightarrow\infty$.
In section 4, we discuss the first law of thermodynamics in the context of dark energy and calculate the various thermodynamical quantities of RT type emergent spacetime with cosmological constant $\L$. The last section is the conclusion of our work.

\section{Brief review of the {\bf k-}essence and Emergent Gravity}
In the {\bf k-}essence theory, the action is \cite{babi1,babi2,babi3,babi4,babi5,gm1,gm2,gm3}
\ben
S_{k}[\phi,g_{\mu\nu}]= \int d^{4}x {\sqrt -g} L(X,\phi)
\label{eq:1}
\een
where $X={1\over 2}g^{\mu\nu}\nabla_{\mu}\phi\nabla_{\nu}\phi$, $g_{\mu\nu}$ is the gravitational metric and $\phi$ is the {\bf k-}essence scalar field.
The energy momentum tensor is
\ben
T_{\mu\nu}\equiv {2\over \sqrt {-g}}{\delta S_{k}\over \delta g^{\mu\nu}}= L_{X}\nabla_{\mu}\phi\nabla_{\nu}\phi - g_{\mu\nu}L
\label{eq:2}
\een
$L_{\mathrm X}= {dL\over dX},~~ L_{\mathrm XX}= {d^{2}L\over dX^{2}},
~~L_{\mathrm\phi}={dL\over d\phi}$ and  
$\nabla_{\mu}$ is the covariant derivative defined with respect to the metric $g_{\mu\nu}$.
The equation of motion is
\ben
-{1\over \sqrt {-g}}{\delta S_{k}\over \delta \phi}= \tilde G^{\mu\nu}\nabla_{\mu}\nabla_{\nu}\phi +2XL_{X\phi}-L_{\phi}=0
\label{eq:3}
\een
where  
\ben
\tilde G^{\mu\nu}\equiv L_{X} g^{\mu\nu} + L_{XX} \nabla ^{\mu}\phi\nabla^{\nu}\phi
\label{eq:4}
\een
and $1+ {2X  L_{XX}\over L_{X}} > 0$.

 The emergent gravity metric $\tilde G_{\mu\nu}$ is not conformally equivalent to the gravitational metric $g_{\mu\nu}$ for the non-trivial configurations of the {\bf k-}essence field $\phi$, $\partial_{\mu}\phi\neq 0$ (for a scalar field,$\nabla_{\mu}\phi\equiv \partial_{\mu}\phi$ ), hence for the $k-$essence theory, the
characteristics are do not coincide with those ones for canonical scalar field the Lagrangian of which depends linearly on the kinetic term $X$. So this {\bf k-}essence field $\phi$ has properties different from canonical scalar fields defined with $g_{\mu\nu}$ and  the local causal structure for this field is different from those one defined by metric $g^{\mu\nu}$.

Carrying out the conformal transformation
$G^{\mu\nu}\equiv {c_{s}\over L_{x}^{2}}\tilde G^{\mu\nu}$, with
$c_s^{2}(X,\phi)\equiv{(1+2X{L_{XX}\over L_{X}})^{-1}}\equiv sound ~ speed $ we have the inverse metric of $G^{\mu\nu}$ is \cite{babi1,babi2,babi3,babi4,babi5}  
\ben G_{\mu\nu}={L_{X}\over c_{s}}[g_{\mu\nu}-{c_{s}^{2}}{L_{XX}\over L_{X}}\nabla_{\mu}\phi\nabla_{\nu}\phi] 
\label{eq:5}
\een
A further conformal transformation \cite{gm1,gm2} $\bar G_{\mu\nu}\equiv {c_{s}\over L_{X}}G_{\mu\nu}$ gives
\ben \bar G_{\mu\nu}
={g_{\mu\nu}-{{L_{XX}}\over {L_{X}+2XL_{XX}}}\nabla_{\mu}\phi\nabla_{\nu}\phi}
\label{eq:6}
\een	
Note that here always have $L_{X}\neq 0$ for the sound speed 
$c_{s}^{2}$ to be positive definite and only 
then equations $(1)-(4)$ will be physically meaningful. 
This can be seen as follows.   $L_{X}=0$ means that $L$ independent of  
$X$ so that in equation (\ref{eq:1}),  $L(X,\phi)\equiv L(\phi)$.
{\it Then  $L$ becomes pure potential but $k-$essence fields correspond to such type of lagrangians where the kinetic energy dominates over the potential energy.} Also the very concept of minimal coupling of $\phi$ to $g_{\mu\nu}$ becomes redundant and 
equation (\ref{eq:1}) meaningless and equations (\ref{eq:4}-\ref{eq:6}) ambiguous.

Again, if we consider $L$ is not an explicit function of $\phi$
then the equation of motion $(3)$ is reduces to:
\ben
-{1\over \sqrt {-g}}{\delta S_{k}\over \delta \phi}
= \bar G^{\mu\nu}\nabla_{\mu}\nabla_{\nu}\phi=0
\label{eq:7}
\een
We shall take the Lagrangian as $L=L(X)=1-V\sqrt{1-2X}$ with $V$ is a constant. 
This is a particular type of the Dirac-Born-Infeld (DBI) \cite{born1,born2,born3} lagrangian
\ben
L(X,\phi)= 1-V(\phi)\sqrt{1-2X}
\label{eq:8}
\een
for $V(\phi)=V=constant$~~and~~$kinetic ~ energy ~ of~\phi>>V$ i.e.$(\dot\phi)^{2}>>V$. This is typical for the {\bf k-}essence field where the kinetic energy dominates over the potential energy.
Then $c_{s}^{2}(X,\phi)=1-2X$.
For scalar fields $\nabla_{\mu}\phi=\partial_{\mu}\phi$. Thus (\ref{eq:6}) becomes
\ben
\bar G_{\mu\nu}= g_{\mu\nu} - \partial _{\mu}\phi\partial_{\nu}\phi
\label{eq:9}
\een

\section{Mapping on to the  Robinson-Trautman type metric with cosmological constant $\L$ when background metric is Reissner-Nordstrom-de Sitter}
We consider the gravitational metric $g_{\mu\nu}$  to be Reissner-Nordstrom- de Sitter (RNdS) and denote $\partial_{0}\phi\equiv\dot\phi$, $\partial_{r}\phi\equiv\phi '$. The line element of RNdS metric \cite{cai,zhao1,zhao2,jg,hs} is 
\ben
ds^2=f(r)dt^2-\frac{dr^2}{f(r)}-r^2 d\O^2
\label{eq:10}
\een
with
\ben
f(r)=(1-\frac{2M}{r}+\frac{Q^2}{r^2}-\frac{\L}{3}r^2)
\label{eq:11}
\een
where $M$ is the mass of the black hole, $Q$ its charge and $\L$ is the cosmological constant ($\L>0$) and $d\O^2=d\theta^2 + sin^{2}\theta~d\Phi^2$.

Assuming that the {\bf k-}essence field $\phi (r,t)$ is spherically symmetric, one has to obtain from (\ref{eq:9})
\ben
\bar G_{00}= g_{00} - (\partial _{0}\phi)^{2}=f(r)- \dot\phi ^{2};\nonumber\\
 \bar G_{11}= g_{11} - (\partial _{r}\phi)^{2}= -\frac{1}{f(r)} - (\phi ') ^{2}; \nonumber\\
\bar G_{22}= g_{22}=-r^{2};
 \bar G_{33}= g_{33}=-r^{2}sin^{2}\theta;
 \nonumber\\
 \bar G_{01}=\bar G_{10}=-\dot\phi\phi '.
\label{eq:12}
\een
The emergent line element is
\ben
ds^2_{emer}=(f(r)-\dot\phi^2)dt^2-(\frac{1}{f(r)}+(\phi')^2)dr^2\nonumber\\-2\dot\phi\phi'~dtdr-r^2d\O^2.
\label{eq:13}
\een
Now making a coordinate transformation such as \cite{gm1,gm2,wein} :
\ben
d\omega=dt-({{\dot\phi\phi '}\over{ f(r) - \dot\phi ^{2}}})dr.
\label{eq:14}
\een
and we choose
\ben
\dot\phi^2=(\phi')^2(f(r))^2
\label{eq:15}
\een
Then the line element (\ref{eq:13}) becomes
\ben
ds^2_{emer}=(f(r)-\dot\phi^2)d\o^2-\frac{dr^2}{(f(r)-\dot\phi^2)}-r^2d\O^2~~
\label{eq:16}
\een

Let us assume a solution to (\ref{eq:15}) of the form 
$\phi(r,t)=\phi_{1}(r)+\phi_{2}(t)$.
Then (\ref{eq:15}) reduces to 
\ben
\dot\phi_{2}^{2}=(\phi_{1}')^{2}(1-\frac{2M}{r}+\frac{Q^2}{r^2}-\frac{\L}{3}r^2)^2= K
\label{eq:17}
\een 
$K(\neq 0)$ is a constant ($K\neq 0$ means {\bf k-}essence field will have {\it non-zero} kinetic energy).
From (\ref{eq:17}) we get
$\dot \phi_{2}(t)=\sqrt{K}$ and $\phi_{1}'(r)=\frac{\sqrt{K}}{f(r)}$.
So the solution of (\ref{eq:15}) is

\ben
\phi(r,t)=\phi_{1}(r)+\phi_{2}(t) \nonumber\\
=-\frac{3\s{K}}{2\L}[A~ln(r^2+l_{1}r+m_{1})+B~ln(r^2-l_{1}r+n_{1})]\nonumber\\
-\frac{3\s{K}}{2\L}[\frac{A(2C-l_{1})}{\s{m_{1}-l_{1}^{2}/4}}~tan^{-1}(\frac{r+l_{1}/2}{\s{m_{1}-l_{1}^{2}/4}})\nonumber\\
+\frac{B(2D+l_{1})}{\s{n_{1}-l_{1}^{2}/4}}~tan^{-1}(\frac{r-l_{1}/2}{\s{n_{1}-l_{1}^{2}/4}})]+\s{K}t\nonumber\\
\label{eq:18}
\een
where
\ben
\phi_{1}(r)=-\frac{3\s{K}}{\L}\int{\frac{r^2dr}{r^4-\frac{3}{\L}r^2+\frac{6M}{\L}r-\frac{3Q^2}{\L}}}\nonumber\\
=-\frac{3\s{K}}{\L}\int{\frac{r^2dr}{(r^2+l_{1}r+m_{1})(r^2-l_{1}r+n_{1})}}\nonumber\\
=-\frac{3\s{K}}{\L}[\int{\frac{Ar+C}{(r^2+l_{1}r+m_{1})}dr}+\int{\frac{Br+D}{(r^2-l_{1}r+n_{1})}dr}]\nonumber\\
=-\frac{3\s{K}}{2\L}[A~ln(r^2+l_{1}r+m_{1})+B~ln(r^2-l_{1}r+n_{1})]\nonumber\\
-\frac{3\s{K}}{2\L}[\frac{A(2C-l_{1})}{\s{m_{1}-l_{1}^{2}/4}}~tan^{-1}(\frac{r+l_{1}/2}{\s{m_{1}-l_{1}^{2}/4}})\nonumber\\
+\frac{B(2D+l_{1})}{\s{n_{1}-l_{1}^{2}/4}}~tan^{-1}(\frac{r-l_{1}/2}{\s{n_{1}-l_{1}^{2}/4}})]\nonumber\\
\label{eq:19}
\een
and 
\ben
\phi_{2}(t)=\s{K}t.
\label{eq:20}
\een
Now we clarify the parameters of the above equation (\ref{eq:18})
$A=-\frac{l_{1}(m_{1}+n_{1})}{N}$, $B=\frac{l_{1}(m_{1}+n_{1})}{N}$,
$C=\frac{n_{1}(n_{1}-m_{1})}{N}$, $D=-\frac{m_{1}(n_{1}-m_{1})}{N}$,
$N=2l_{1}^2(m_{1}+n_{1})+(m_{1}-n_{1})^2$,
$l_{1}=[(\frac{E_{1}+\s{E_{1}^2-4D_{1}^3}}{2})^{1/3}+(\frac{E_{1}-\s{E_{1}^2-4D_{1}^3}}{2})^{1/3}+\frac{2}{\L}]^{1/2}$, 
$m_{1}=\frac{1}{2\L l_{1}}(\L l_{1}^3-3l_{1}-6M)~,~n_{1}=\frac{1}{2\L l_{1}}(\L l_{1}^3-3l_{1}+6M)$,
$E_{1}=\frac{1}{\L^3}[(36M^2-24Q^2)\L-2]~,~D_{1}=\frac{1}{\L^2}(1-4Q^{2}\L)$

Applying the equation (\ref{eq:17}) in equation (\ref{eq:16}) we get the emergent line element
\ben
ds^2_{emer}=(1-\frac{2M}{r}+\frac{Q^2}{r^2}-\frac{\L}{3}r^2-K)d\o^2 \nonumber\\
-\frac{dr^2}{(1-\frac{2M}{r}+\frac{Q^2}{r^2}-\frac{\L}{3}r^2-K)}-r^{2}d\O^2\nonumber\\
=(\epsilon-\frac{2M}{r}+\frac{Q^2}{r^2}-\frac{\L}{3}r^2)d\o^2\nonumber\\
-\frac{dr^2}{(\epsilon-\frac{2M}{r}+\frac{Q^2}{r^2}-\frac{\L}{3}r^2)}
-r^{2}d\O^2
\label{eq:21}
\een
with $\epsilon=(1-K)=constant$.

Now we are using the Eddington-Finkelstein coordinates $(v,r)$ or 
$(u,r)$ i.e., {\it introducing advanced and retarded null coordinates}
\ben
v=\omega+r^{*}~~;~~u=\omega-r^{*}~;~\nonumber\\
 dr^{*}=\frac{dr}{(\epsilon-\frac{2M}{r}+\frac{Q^2}{r^2}-\frac{\L}{3}r^2)}
\label{eq:22}
\een
Then from (\ref{eq:21}) we get
\ben
ds^2_{emer}=(\e-\frac{2M}{r}+\frac{Q^2}{r^2}-\frac{\L}{3}r^2)dv^{2}-2dvdr-r^{2}d\O^2\nonumber\\
\label{eq:23}
\een

For the solution (\ref{eq:18}) of $\phi(r,t)$, the line element (\ref{eq:21}) and (\ref{eq:23}) are analogous to the {\it Robinson-Trautman} (RT) type metric with cosmological constant $\L$ {\it (\cite{jg}, Chapter-9, page no.-166, Chapter-19 and \cite{hs}, Chapter-28)}. The $\L$ parameter comes from the background RN-de Sitter gravitational metric. This RT type spacetime contains surfaces of constant positive, zero or negative Gaussian curvature according to whether $\epsilon=+1,0,-1$ respectively.

In our case $\epsilon \neq +1$ since if $\epsilon=1$ then $K=\dot\phi_{2}^{2}=0$ which implies  dark energy is absent but in the {\bf k-}essence theory, dark energy density (in unit of critical density)  i.e. kinetic energy of the {\bf k-}essence scalar field can not be zero. Also, $\epsilon\neq -1$ since if $\epsilon=-1$ then $K=2$ but the total energy density 
cannot exceed unity ($\Omega_{matter} +\Omega_{radiation} +\Omega_{dark energy}= 1$). Although $\epsilon=1$ is a valid solution in every spherically symmetric mass distribution starting from earth, star, etc. even in the presence of the cosmological constant.
So, for the above reasons, the RT type emergent solution of the Einstein equation provided us $\epsilon=0$ i.e., $K=1$ which implies 
$\Omega_{dark energy}= 1$ and the other energy densities are taken to be zero. 
So we can consider this is the special case where the emergent spacetime is only dark energy dominated. This is one possibility for satisfying the Einstein equation in the context of emergent spacetime. 
Here we have shown that the total energy density cannot exceed unity by mentioning the relation $\Omega_{matter} +\Omega_{radiation} +\Omega_{dark energy}= 1$ \cite{gm2} only to get the permissible values of $\epsilon$ i.e., $K$ and most importantly  we are not relating it with Friedmann-Lemaitre-Robertson-Walker (FLRW) metric since we have not considered FLRW spacetime as a background.

Note that dark energy density (K) i.e., K.E. of the {\bf k-}essence scalar field is the essential property of the {\bf k-}essence theory which can not be redefined with the radial coordinate $r$ to absorb the $\epsilon$. Also,
the case K=1 of (\ref{eq:21}) does not seem to have a 
Newtonian limit \cite{schutz}, which makes it unsuitable for describing astrophysical objects. However, this may not 
be suitable as an astrophysical object but still,  it is a consistent solution to Einstein's equation.

Also, mention that the emergent line element (\ref{eq:21}) or (\ref{eq:23}) is not the dRGT type massive gravity \cite{rham1,rham2} line element.
In a massive-gravity model \cite{buri,ghosh}, all the physical parameters depend on graviton mass $m_{g}$ and in the line element, present the linear term of $r$ in addition to the cosmological constant. But in our case, there is no linear term of $r$ is present in the line element (\ref{eq:23}). Here we are mapping the emergent gravity spacetime in the context of dark energy for a particular type of {\bf k-}essence scalar field (\ref{eq:18}) with the RT type spacetime only.

Note that the solution $\phi(r,t)$ (\ref{eq:18}) is satisfied emergent gravity equation of motion (\ref{eq:7}) at $r\rightarrow\infty$: 
$\bar G^{00}\partial_{0}^{2}\phi_{\mathrm 2} 
+ [\bar G^{11}(\partial_{1}^{2}\phi_{\mathrm 1} 
-\Gamma_{11}^{1}\partial_{1}\phi_{\mathrm 1})]
+\bar G^{01}\nabla_{0}\nabla_{1}\phi
+\bar G^{10}\nabla_{1}\nabla_{0}\phi
+\bar G^{22}\nabla_{2}\nabla_{2}\phi
+\bar G^{33}\nabla_{3}\nabla_{3}\phi= 0$.
The first term vanishes since $\phi_{2}(t)$ linear in $t$ and second term within third bracket vanish
at $r\rightarrow\infty$ because 
$2nd~term =-\frac{\s{K}}{2}\frac{(3\epsilon r^2-6Mr+3Q^2-\L r^4)(6\L r^4-18Mr+18Q^2)}{3r(\L r^4-3r^2+6Mr-3Q^2)^{2}} $ and
 the third and fourth terms vanish because $\bar G^{01}=\bar G^{10}=0$
 and the last two terms vanish since $\phi$ is not dependent on $\theta$ and $\Phi$ in this case.

\section{Thermodynamics of Robinson-Trautman type metric with Cosmological constant $\L$}
We rewrite the RT metric (\ref{eq:21}) with cosmological constant $\L$  as
\ben
ds^2_{RT}=F(r)d\o^2-\frac{dr^2}{F(r)}-r^{2}d\O^2
\label{eq:24}
\een
where
\ben
F(r)=(\epsilon-\frac{2M}{r}+\frac{Q^2}{r^2}-\frac{\L}{3}r^2).
\label{eq:25}
\een
At the horizon $F(r)=0$, which admits the quartic equation of $r$. The roots are
\ben
r=\frac{-l\pm\s{l^{2}-4m}}{2}~,~\frac{l\pm\s{l^{2}-4n}}{2}
\label{eq:26}
\een
where
$l=[(\frac{E+\s{E^2-4W^3}}{2})^{1/3}+(\frac{E-\s{E^2-4W^3}}{2})^{1/3}+\frac{2\e}{\L}]^{1/2}$~;~
$m=\frac{1}{2}(\frac{-3\e}{\L}+l^2-\frac{6M}{\L l})~,~n=\frac{1}{2}(\frac{-3\e}{\L}+l^2+\frac{6M}{\L l})$~;~ $E=\frac{1}{\L^3}[(36M^2-24Q^2\e)\L-2\e^3]~;~W=\frac{1}{\L^2}(\e^2-4Q^{2}\L)$.

Depending upon the values of parameters $M,Q,\L~and~\e$ we can denote the roots are $r_{cd}, r_{+d}, r_{-d}, r_{4d}$ in the presence of dark energy. The order of roots are $r_{cd}>r_{+d}>r_{-d}>r_{4d}$ where $r_{4d}<0$. The root $r_{cd}$ denotes the location of the cosmological event horizon, $r_{+d}$ denotes the location of the black hole event (outer)  horizon and $r_{-d}$ denotes the location of the Cauchy (inner)  horizon  of this black hole spacetime (\ref{eq:24}) \cite{zhao1,zhao2}. Note that $r_{4d}$ does not correspond to a physical horizon in this spacetime.

The equations 
\ben
F(r_{+d})=0=Q^2-2Mr_{+d}+r_{+d}^2(\e-\frac{\L}{3}r_{+d}^2)
\label{eq:27}
\een
and
\ben
F(r_{cd})=0=Q^2-2Mr_{cd}+r_{cd}^2(\e-\frac{\L}{3}r_{cd}^2)
\label{eq:28}
\een
are rearranged in terms of $r_{cd}~and~r_{+d}$ as
\ben
Q^2=r_{cd}r_{+d}[\e-\frac{\L}{3}(r_{cd}^2+r_{cd}r_{+d}+r_{+d}^2)]
\label{eq:29}
\een
and
\ben
2M=(r_{+d}+r_{cd})[\e-\frac{\L}{3}(r_{cd}^2+r_{+d}^2)].
\label{eq:30}
\een
The surface gravity on the black hole and the cosmological horizons are
\ben
\k_{+d}=\frac{1}{2}\frac{dF}{dr}\arrowvert_{r=r_{+d}}
=\frac{1}{2r_{+d}^3}[2Mr_{+d}-2Q^2-2\frac{\L}{3}r_{+d}^4]\nonumber\\
=\frac{1}{2r_{+d}^3}[r_{+d}^{2}\e-Q^2-\L r_{+d}^4]\nonumber\\
=\frac{r_{+d}-r_{cd}}{2r_{+d}^2}[\e-\frac{(r_{cd}^2+2r_{cd}r_{+d}+3r_{+d}^2)(\e r_{cd}r_{+d}-Q^2)}{r_{cd}r_{+d}(r_{+d}^2+r_{cd}r_{+d}+r_{cd}^2)}]\nonumber\\
=\frac{r_{+d}-r_{cd}}{2r_{+d}^2}[\e-\frac{\L}{3}(r_{cd}^2+2r_{+d}r_{cd}+3r_{+d}^2)]\nonumber\\
\label{eq:31}
\een
and
\ben
\k_{cd}=\frac{1}{2}\frac{dF}{dr}\arrowvert_{r=r_{cd}}
=\frac{1}{2r_{cd}^3}[2Mr_{cd}-2Q^2-2\frac{\L}{3}r_{cd}^4]\nonumber\\
=\frac{1}{2r_{cd}^3}[r_{cd}^{2}\e-Q^2-\L r_{cd}^4]\nonumber\\
=\frac{r_{cd}-r_{+d}}{2r_{cd}^2}[\e-\frac{(r_{+d}^2+2r_{cd}r_{+d}+3r_{cd}^2)(\e r_{cd}r_{+d}-Q^2)}{r_{cd}r_{+d}(r_{+d}^2+r_{cd}r_{+d}+r_{cd}^2)}]\nonumber\\
=\frac{r_{cd}-r_{+d}}{2r_{cd}^2}[\e-\frac{\L}{3}(r_{+d}^2+2r_{+d}r_{cd}+3r_{cd}^2)]\nonumber\\
\label{eq:32}
\een

The first law of thermodynamics for the black hole horizon and the cosmological horizon are expressed as \cite{zhao1,dolan,seki,gibbon1,gibbon2,li,kubiznak}:
\ben
\d M=\frac{\k_{+d}}{2\pi}\d S_{+d}+\varphi_{+d}\d Q+V_{+d}\d P
\label{eq:33}
\een
and
\ben
\d M=\frac{\k_{cd}}{2\pi}\d S_{cd}+\varphi_{cd}\d Q+V_{cd}\d P
\label{eq:34}
\een
where
$S_{+d}=\pi r_{+d}^2$, $S_{cd}=\pi r_{cd}^2$, $\varphi_{+d}=Q/r_{+d}$, $\varphi_{cd}=-(Q/r_{cd})$, $V_{+d}=\frac{4\pi}{3}r_{+d}^3$,
$V_{cd}=\frac{4\pi}{3}r_{cd}^3$ and $P=-\frac{\L}{8\pi}$.
This law means that total energy of the black hole system is
conserved.

In \cite{zhao1,zhao2,ren} obtained that the outgoing rate of the
charged de Sitter spacetime which radiates particles with
energy $\o$ is $\Gamma=e^{(\Delta S_{+d}+\Delta S_{cd})}$
where $\Delta S_{+d}$ and $\Delta S_{cd}$ are Bekenstein-Hawking entropy difference in the presence of dark energy corresponding to the black hole horizon and the cosmological horizon after the charged de Sitter spacetime radiates particles with energy $\o$. Therefore, in this context, we can say that the thermodynamic entropy of the RT type spacetime in the presence of dark energy is the sum of
the black hole horizon entropy and the cosmological horizon
entropy (as in \cite{zhao1,zhao2}):
\ben
S=S_{+d}+S_{cd}.
\label{eq:35}
\een
substituting the values of $\d S_{+d}$ and $\d S_{cd}$
from equations (\ref{eq:33} and \ref{eq:34}) in equation (\ref{eq:35}) we get
\ben
dS=2\pi(\frac{1}{\k_{+d}}+\frac{1}{\k_{cd}})dM-2\pi(\frac{\varphi_{+d}}{\k_{+d}}+\frac{\varphi_{cd}}{\k_{cd}})dQ\nonumber\\-2\pi(\frac{V_{+d}}{\k_{+d}}+\frac{V_{cd}}{\k_{cd}})dP=2\pi(\frac{1}{\k_{+d}}+\frac{1}{\k_{cd}})dM\nonumber\\-2\pi(\frac{\varphi_{+d}}{\k_{+d}}+\frac{\varphi_{cd}}{\k_{cd}})dQ+\frac{\pi}{3}(\frac{r_{+d}^3}{\k_{+d}}+\frac{r_{cd}^3}{\k_{cd}})d\L.
\label{eq:36}
\een

For the sake of simplicity, we consider the uncharged case, then the above equation (\ref{eq:36}) become
\ben
dS=2\pi(\frac{1}{\k_{+d}}+\frac{1}{\k_{cd}})dM+\frac{\pi}{3}(\frac{r_{+d}^3}{\k_{+d}}+\frac{r_{cd}^3}{\k_{cd}})d\L.
\label{eq:37}
\een
Differentiating equations (\ref{eq:27}) and (\ref{eq:28}) and using equations (\ref{eq:31}) and (\ref{eq:32}) we get
\ben
dr_{+d}=\frac{dM}{r_{+d}\k_{+d}}+\frac{(r_{+d}^{3}/3)}{2r_{+d}\k_{+d}}d\L
\label{eq:38}
\een
and
\ben
dr_{cd}=\frac{dM}{r_{cd}\k_{cd}}+\frac{(r_{cd}^{3}/3)}{2r_{cd}\k_{cd}}d\L.
\label{eq:39}
\een

The thermodynamic volume \cite{zhao1,dolan,urano} of RT type spacetime (\ref{eq:24}) is given as
\ben
V=\frac{4\pi}{3}(r_{cd}^{3}-r_{+d}^3).
\label{eq:40}
\een

Now substituting the values of $dr_{+d}$ and $dr_{cd}$ in (\ref{eq:40}) we get
\ben
dV=4\pi(\frac{r_{cd}}{\k_{cd}}-\frac{r_{+d}}{\k_{+d}})dM+\frac{2\pi}{3}(\frac{r_{cd}^4}{\k_{cd}}-\frac{r_{+d}^4}{\k_{+d}})d\L.
\label{eq:41}
\een
Using equations (\ref{eq:37}) and (\ref{eq:41}), one can derive the thermodynamic equation \cite{zhao1,urano}
\ben
dM=T_{eff}^{d}dS-P_{eff}^{d}dV.
\label{eq:42}
\een
The effective temperature $T_{eff}^{d}$ is
\ben
T_{eff}^{d}=[\frac{(y^4+y^3-2y^2+y+1)}{4\pi r_{cd}y(y+1)(y^2+y+1)}]\e\nonumber\\-\frac{Q^2}{4\pi}[\frac{(y^6+y^5+y^4-2y^3+y^2+y+1)}{ r_{cd}^{3}y^3(y+1)(y^2+y+1)}]
\label{eq:43}
\een
and the effective pressure $P_{eff}^{d}$ is
\ben
P_{eff}^{d}=[\frac{(1-y)(1+3y+3y^2+3y^3+y^4)}{8\pi r_{cd}^{2}y(1+y)(1+y+y^2)^{2}}]\e\nonumber\\
-\frac{Q^2}{8\pi}[\frac{(1-y)(1+2y+3y^2-3y^5-2y^6-y^7)}{ r_{cd}^{4}y^3(1+y)(1-y^3)(1+y+y^2)}]
\label{eq:44}
\een
where $y=\frac{r_{+d}}{r_{cd}}$ and $0<y<1$.

Because of the temperature and pressure are positive definite for ordinary laboratory systems, it is natural that the thermodynamical stability should  express appropriately and it should satisfy the following requirement:
$T_{eff}^{d}>0$ and $P_{eff}^{d}>0$. 

But in our case, we are restricting the value of $\e$ is $\e=0 ~i.e.~K=1$, which implies the following situations:

a) For $Q\neq 0$, it is evident from (\ref{eq:43}) and (\ref{eq:44}) that the effective temperature and pressure are negative. {\it Negativity of pressure indicates that the universe is dark energy dominated whereas negative temperature stands for a thermodynamically unstable system.} So in this situation, for the restriction of the dark energy density limiting to unity, RT type spacetime (\ref{eq:24}) is thermodynamically unstable though this type of spacetime is a valid solutions of Einstein's field equation. We can describe this type of situation as {\it per Ref. \cite{reif}: thermodynamically unstable situation is physically plausible only when we consider spin degrees of freedom  of a system without taking translational degrees of freedom (i.e., coordinates and momenta of the system) into account.
Based on this consideration the number of possible spin states, at first, increase with energy leading to a maximum and finally decreases again. This feature indicates that the absolute spin temperatures may be negative as well as positive. So we can infer that the RT type black hole (\ref{eq:24}) can be projected as a physical reality only by considering spin degrees of freedom an exceptional condition.}

b) Another condition: if $\e=0$ and $Q=0$ then we see that from (\ref{eq:43}) and (\ref{eq:44}) the effective temperature and pressure are both zero. Zero temperature means
that the RT type black hole (\ref{eq:24}) does not radiate and zero pressure means the volume-thermodynamic system becomes area-thermodynamic since horizons have nonzero area.

Now we are considering the problem when the black hole event horizon and the cosmological event horizon are independent of each other i.e., the two horizons are coincide \cite{zhao1}, then, 
\ben
r_{+d/cd}^{2}=\frac{\e\pm \s{\e^2-4Q^{2}\L}}{2\L}\equiv r_{0d}^{2}.
\label{eq:45}
\een
From above equation (\ref{eq:45}) we can see that $r_{0d}^{2}=\frac{iQ}{\s{\L}}$ i.e., imaginary horizon since in our case the dark energy density is to be unity ($K=1$) i.e., $\e=0$. Again from (\ref{eq:31}) and (\ref{eq:32}) we see that the surface gravity of the two horizons are zero as two horizons are coinciding. Thus the temperatures of both horizons are zero in the presence of dark energy when two horizons are coincided. However, both horizons have the nonzero area, which means that the entropy for the two horizons should not be zero.

\section{Conclusion}

We have investigated the thermodynamics of a black hole in the context of dark energy in an emergent gravity scenario using {\bf k-}essence model for Reissner -Nordstrom - de Sitter background. The emergent gravity spacetime has to be conformally invariant with the Robinson -Trautman type spacetime with cosmological constant $\L$.
The evaluated thermodynamical quantities like surface gravity, area law, temperature, pressure etc.  are different from the usual ones.
As our restriction of the emergent spacetime (23), taken to be the dark energy density i.e., the kinetic energy of the {\bf k-}essence scalar field is unity, the RT type emergent black hole does not radiate when $Q=0$ and {\it when $Q\neq 0$ the system possess thermodynamically unstable equilibrium.}{\it So we can say that the RT type emergent black hole (\ref{eq:23}) is thermodynamically unstable in the presence of dark energy and which can be projected as a physical reality only by considering spin degrees of freedom an exceptional condition.} Also, for this restriction in the RT type emergent metric (23), we conclude that this type of spacetime is only dark energy dominated as a special consideration since it is a valid solution of the Einstein equation.

\vspace{0.2in}

{\bf Acknowledgement:} 
The authors thanks to Dr. Pradipta Panchadhyayee, Associate Professor, Department of Physics, Prabhat Kumar College, Contai for his valuable suggestions. The authors would like to thank the referees for illuminating suggestions to improve the manuscript.


\vspace{0.3in}





\begin{thebibliography}{99}
\bibitem{babi1}
M.Visser,C.Barcelo and S.Liberati, Gen.Rel.Grav. {\bf 34} 1719 (2002)
\bibitem{babi2}
E.Babichev, V.Mukhanov and A.Vikman, JHEP {\bf 09}, 061 (2006)
\bibitem{babi3}
E.Babichev,M.Mukhanov and A.Vikman, JHEP {\bf 0802} 101 (2008)
\bibitem{babi4}
E.Babichev,M.Mukhanov and A.Vikman, WSPC-Proceedings, February 1, 2008
\bibitem{babi5}
Alexander Vikman, {\it K-essence: Cosmology, causality and Emergent Geometry},
 Dissertation an der Fakultat fur Physik,Arnold Sommerfeld Center for Theoretical Physics, 
der Ludwig-Maximilians-Universitat Munchen, Munchen, den 29.08.2007.

\bibitem{gm1}
D.Gangopadhyay and Goutam Manna, Euro.Phys.Lett. {\bf 100} 49001 (2012).
\bibitem{gm2}
Goutam Manna and Debashis Gangopadhyay, Eur. Phys. J. C {\bf 74} 2811 (2014).
\bibitem{gm3}
Goutam Manna and Bivash Majumder, Eur. Phys. J. C {\bf 79}, 553, (2019).

\bibitem{scherrer1}
R.J. Scherrer, Phys.Rev.Lett.{\bf 93} 011301 (2004)
\bibitem{scherrer2}
L.P.Chimento, Phys.Rev.{\bf D69} 123517 (2004)
\bibitem{scherrer3}
D.Gangopadhyay and S. Mukherjee, Phys. Lett.{\bf B665} 121 (2008).


\bibitem{gor1}
V.Gorini,A.Kamenschik and U.Moschella, Phys.Rev. {\bf D67} 063509 (2003)
\bibitem{gor2}
V.Gorini,A.Kamenschik and U.Moschella and V.Pasquier ,arXiv:gr-qc/0403062 (2004)
\bibitem{gor3}
L.Rizzi,S.Cacciatori,V.Gorini,A.Kamenschik and O.F.Piatella, Phys.Rev {\bf D82} 027301 (2010)
\bibitem{gor4}
A.Y.Kamenschik,A.Tronconi and G.Venturi, Phys.Lett. {\bf B702} 191 (2011)
\bibitem{gor5}
C.Armendariz-Picon, T.Damour and V.Mukhanov, Phys.Lett.{\bf B458} 209 (1999)
\bibitem{gor6}
C.Armendariz-Picon, V.Mukhanov and P.J.Steinhardt, Phys.Rev.{\bf D63} 103510 (2001)
\bibitem{gor7}
T.Chiba, T.Okabe and M.Yamaguchi, Phys.Rev.{\bf D62} 023511 (2000)
\bibitem{gor8}
C.Armendariz-Picon and E.A.Lim, JCAP {\bf 0508} 007 (2005)
\bibitem{gor9}
N.Arkani-Hamed, H.C.Cheng,M.A.Luty and S.Mukohyama, JHEP {\bf 05} 074 (2004)
\bibitem{gor10}
N.Arkani-Hamed, P.Creminelli,S.Mukohyama and M.Zaldarriaga, JCAP {\bf 0404} 001 (2004)
\bibitem{gor11}
R.R.Caldwell, Phys.Lett.{\bf B545} 23 (2002)
\bibitem{gor12}
J.Callan, G.Curtis and J.M.Maldacena, Nucl.Phys.{\bf B513} 198 (1998)
\bibitem{gor13}
A.D.Rendall, Class.Quant.Grav.{\bf 23} 1557 (2006).
\bibitem{gor14}
G.W.Gibbons,  Nucl.Phys.{\bf B514} 603 (1998)
\bibitem{gor15}
G.W.Gibbons, Rev.Mex.Fis.{\bf 49S1} 19 (2003)


\bibitem{born1}
M.Born and L.Infeld,Proc.Roy.Soc.Lond {\bf A144}(1934) 425
\bibitem{born2}
W.Heisenberg,  Zeitschrift fur Physik A Hadrons and Nuclei {\bf 113} no.1-2 (1939)
\bibitem{born3}
P.A.M.Dirac, Royal Society of London Proceedings Series A {\bf 268} (1962) 57.







\bibitem{haw1}
S. Hawking, Phys. Rev. Letters {\bf 26}, 1344 (1971)
\bibitem{haw2}
L. Smarr, Phys. Rev. Lett. 30, {\bf 71} (1973)
\bibitem{haw3}
J. Bardeen, B. Carter and S. Hawking, Comm. Math. Phys. {\bf 31}, 161 (1973)
\bibitem{haw4}
S. Hawking, Nature (London) 248, {\bf 30} (1974)
\bibitem{haw5}
S. Hawking, Commun. Math. Phys. {\bf 43}, 199 (1975)
\bibitem{haw6}
J. Bekenstein, Phys. Rev. {\bf D7}, 2333 (1973); Phys. Rev. {\bf D9}, 3292 (1974)
\bibitem{haw7}
G. Gibbons and S. Hawking, Phys. Rev. {\bf D15}, 2752 (1977)
\bibitem{haw8}
S.W. Hawking, G.T. Horowitz and S.F. Ross, Phys. Rev. {\bf D51}, 4302 (1995)
\bibitem{haw9}
Maulik K. Parikh and Frank Wilczek, Phys.Rev.Lett.{\bf 85}, 5042 (2000)
\bibitem{haw10}
K.Murata and J.Soda, Phys.Rev.{\bf D74}, 044018 (2006)
\bibitem{haw11}
R.Kerner and R.B.Mann, Class.Quant.Grav. {\bf 25}, 095014 (2008)
\bibitem{haw12}
P.Mitra, Phys.Lett. {\bf B648}, 240 (2007)
\bibitem{haw13}
Bhramar Chatterjee, A.Ghosh and P.Mitra, Phys.Lett. {\bf B661}, 307 (2008)
\bibitem{haw14}
Bhramar Chatterjee and P.Mitra, Phys.Lett. {\bf B675}, 640 (2008)
\bibitem{haw15}
P.Mitra, {\it Black Hole Entropy}, [arXiv: 0902.2055].





 \bibitem{cai}
R.G. Cai,  Nuclear Physics B, {\bf 628, no.1-2}, 375-386, (2002).
\bibitem{zhao1}
Ren Zhao, Mengsen Ma, Huihua Zhao, and Lichun Zhang, Advances in High Energy Physics, Volume 2014, Article ID 124854, Hindawi Publishing Corporation, 2014.
\bibitem{zhao2}
Ren Zhao, Li-Chun Zhang and Huai-Fan Li, Gen. Relativ. Gravit. {\bf 42}, 975, (2010).


\bibitem{dolan}
B. P. Dolan, D. Kastor, D. Kubiznak, R. B. Mann, and J. Traschen,
Physical Review D, {\bf 87}, 104017, (2013).
\bibitem{seki}
Y. Sekiwa, Physical Review D, {\bf 73}, 084009, (2006).
\bibitem{gibbon1}
G. W. Gibbons, H. Lü, D. N. Page, and C. N. Pope,
Physical Review Letters, {\bf 93}, 171102, (2004).
\bibitem{gibbon2}
G. W. Gibbons, H. Lü, D. N. Page, and C. N. Pope, Journal of Geometry
and Physics, {\bf 53}, 49–73, (2005).
\bibitem{ren}
Z. Ren, Z. Lichun, L. Huaifan, and W. Yueqin,  European
Physical Journal C, {\bf 65}, 289–293, (2009).
\bibitem{urano}
M. Urano and A. Tomimatsu, Classical and Quantum
Gravity, {\bf 26}, 105010, (2009).
\bibitem{li}
Huai-Fan Li, Meng-Sen Ma, and Ya-Qin Ma, Modern Physics Letters A
{\bf 32}, No. 2, 1750017, (2017).
\bibitem{kubiznak}
David Kubiznak and Fil Simovic, Class. Quantum Grav., {\bf 33}, 245001, (2016).






\bibitem{buri}
Piyabut Burikham et. al. Physical Review {\bf D96}, 124001 (2017).
\bibitem{ghosh}
Suchant G. Ghosh et. al., Eur. Phys. J. C {\bf 76}:119, (2016) 
\bibitem{rham1}
Claudia de Rham and Gregory Gabadadze, Physical Review {\bf D82}, 044020 (2010)
\bibitem{rham2}
Claudia de Rham et. al., PRL, {\bf 106}, 231101 (2011)


\bibitem{jg}
Jerry B. Griffiths and Jiri Podolsky, {\it Exact Space-Times in Einstein's General Relativity}, Cambridge Monographs on Mathematical Physics, Cambridge University Press, (2009).
\bibitem{hs}
Hans Stephani, D. Kramer, M. Maccallum, C. Hoenselaers and E. Herlt, {\it Exact Solutions of Einstein's Field Equations}, Cambridge Monographs on Mathematical Physics, Cambridge University Press, (2003)


\bibitem{wein}
S. Weinberg, {\it Gravitation and Cosmology}, Wiley Student Edition, John Wiley and Sons (Asia) Pte. Ltd., 2004

\bibitem{schutz}
Bernard F.Schutz, {\it A first course in General Relativity, Ch.8, section 4}, 
Cambridge University Press, (1985).

\bibitem{reif}
F. Reif, {\it Fundamentals of Statistical and Thermal Physics, Ch.3, section 3.5, page 105}, Levant, Indian Edition, (2010)





\end{thebibliography}
\end{document}